\documentclass[12pt]{article}
\usepackage{geometry}                
\usepackage{graphicx}
\usepackage{amssymb}
\usepackage{amsmath}
\usepackage{amsfonts}
\usepackage{graphicx}
\usepackage{epstopdf}
\usepackage{verbatim}
\usepackage{color}
\usepackage{cite}
\usepackage[normalem]{ulem}
\usepackage{float}

\newcommand{\be}{\begin{equation}}
\newcommand{\ee}{\end{equation}}
\newcommand{\bea}{\begin{eqnarray}}
\newcommand{\eea}{\end{eqnarray}}
\newcommand{\beas}{\begin{eqnarray*}}
\newcommand{\eeas}{\end{eqnarray*}}
\newcommand{\avg}[1]{\left\langle{#1}\right\rangle}

\newcommand{\w}{\vec{w}}

\newcommand{\xk} {{\vec x}_{\tau}}
\newcommand{\wxk} {{\vec w} \cdot \xk }

\newcommand{\cor}{\color{black}}
\newcommand{\cob}{\color{black}}

\title{$L_p$ Regularized Portfolio Optimization}
\date{\today}

\author{Fabio Caccioli,\\ {\em Centre for Risk Studies, Cambridge Judge Business School}\\{\em University of Cambridge, Trumpington St, Cambridge  CB2 1AG , UK}  \and Imre Kondor,\\{\em Parmenides Foundation, Kirchplatz 1, 82049 Pullach, Germany} \and Matteo Marsili,\\ {\em Abdus Salam International Centre for Theoretical Physics,}\\ {\em Strada Costiera 11, 34151 Trieste, Italy}\and Susanne Still,\\ {\em Information and Computer Sciences, University of Hawai`i at}\\ {\em M\={a}noa, 1680 East-West Road, Honolulu 96822, Hawai`i, USA}}

\begin{document}
\bibliographystyle{unsrt}

\maketitle

\begin{abstract} 
Investors who optimize their portfolios under any of the coherent risk measures are naturally led to regularized portfolio optimization when they take into account the impact their trades make on the market. We show here that the impact function determines which regularizer is used. We also show that any regularizer based on the norm $L_p$ with $p>1$ makes the sensitivity of coherent risk measures to estimation error disappear, while regularizers with $p<1$ do not. The $L_1$ norm represents a border case: its ``soft'' implementation does not remove the instability, but rather shifts its locus, whereas its ``hard'' implementation (equivalent to a ban on short selling) eliminates it. We demonstrate these effects on the important special case of Expected Shortfall (ES) that is on its way to becoming the next global regulatory market risk measure. 
\end{abstract}

\section{Introduction}
{\cor Risk minimization for large institutional portfolios suffers from the curse of dimensionality: the number of different assets (the dimension of the measured return vector), $N$, is often comparable to, or even larger than the sample size (the length of the available time series), $T$.
As such, portfolio optimization belongs to the realm of high dimensional statistics \cite{buh11}. Empirical estimates of returns, covariances, etc. are unstable under sample fluctuations.} 

The instability has been identified by \cite{kondor2007, CKM07} as an algorithmic phase transition \cite{ Mezard2009}, occurring at a critical value of the ratio $N/T$ (depending on the risk measure in question). At this critical point the estimation error diverges with a universal exponent, independent of the risk measure, the nature of the underlying fluctuations, or even whether these fluctuations are stationary or GARCH-like \cite{hasszan2007}. 

Expected Shortfall (ES) \cite{Rockafellar,acerbi1} is on its way to becoming the next global regulatory market risk measure \cite{Basel2012, Basel2013}. The sensitivity of ES to estimation error was studied in \cite{kondor2007,CKM07,hasszan2008}, where the critical value of the ratio $N/T$ was determined, at which the estimation error diverges and the risk function becomes unbounded from below. This instability is a common weakness \cite{kondor2008} of all coherent risk measures \cite{Artzner99}. 

Various methods have been proposed to address this problem
\cite{eltongruber,jobson1979,jorion1986,frost_savarino,safePO,Jagannathan2003, LedoitWolf2003, LedoitWolf2004, LedoitWolfHoney, DeMiguel2007, Garlappi2007,Golosnoy2007,Kan2007,frahm_memmel,DeMiguel2009, Brodie2009, Laloux1999,Plerou1999,Laloux2000,Plerou2000,burda,Potters2005}. In \cite{RPO2010} we advocated the systematic use of regularization in portfolio selection to eliminate the instability. 
We illustrated the idea of regularized portfolio optimization (RPO) using the Expected Shortfall as risk measure and the $L_2$ norm as regularizer \cite{RPO2010}. The optimization problem thus obtained was shown \cite{Nu-SVM,Takeda2008,RPO2010}\footnote{{\cor This relationship was first recognized in \cite{Takeda2008}, but the budget constraint was omitted there, which is an important element of the portfolio problem, missing from support vector regression.}} to be closely related to support vector regression{\cor \cite{Nu-SVM}}.

The financial content of regularization is revealed by realizing that regularized portfolio optimization arises naturally when investors take into account the impact their trades make on the market \cite{CSMK}. We demonstrated (again using the specific example of ES) that the $L_2$ norm regularizer corresponds to linear market impact, and that the resulting optimization problem is identical to RPO, eliminating the instability \cite{CSMK}. 

In this paper we show that considering market impact in the investment strategy results in a regularized portfolio optimization {\em for all coherent risk measures}. We identify the various regularizers corresponding to the different impact functions. 
We furthermore prove that the instability of {\em all} coherent measures can be removed by regularizers based on the $L_p$ norm with $p>1$, but not by regularizers with $p<1$. 

The $L_1$ norm sits on the borderline: it may or may not remove the instability, depending on the sample, which means that on average it does not remove, but only shifts it. In order to prevent the instability, the $L_1$ norm regularizer has to be implemented as a ban on short selling, which corresponds to a hard constraint, i.e. imposing infinite penalty on solutions wandering outside a finite domain. 

We illustrate the effect of various regularizers on the estimation error of Expected Shortfall, as it has acquired special significance recently since the Basel Committee proposed it as the next global regulatory market risk measure \cite{Basel2012, Basel2013}. The calculations can be carried through analytically in the limit of large Gaussian portfolios ($N,T\to\infty$, with their ratio fixed) by methods borrowed from the statistical physics of disordered systems.

The paper is organized as follows. In Section \ref{Imp-Reg} we show how market impact considerations lead to regularization for all coherent risk measures, and display the correspondence between the various impact functions and regularizers. Section \ref{L1-inst} analyzes the effect of various $L_p$ regularizers on estimation error, with particular focus on the $L_1$ norm. We start by analyzing a simple toy example: the Maximal Loss risk function (the limiting case of ES) with only two assets and two data points. This serves to provide intuition. We move on to show that the $L_1$ regularizer cannot remove the instability of any of the coherent risk measures, while an $L_p$ with $p>1$ can. Section \ref{L1ES} illustrates the above ideas using the example of a large random portfolio optimized under historical ES, and demonstrates the shift of the instability as a result of $L_1$-regularization. The effect of $L_p$, $p>1$, regularizers is also discussed, with special emphasis on the case $p=3/2$ corresponding to a square root-like market impact that is characteristic of the liquidation of large positions under normal market conditions. A combination of the $L_1$ and $L_2$ norms as a regularizer is also considered in this Section. The last Section is a short summary, followed by the Appendix, where the technical details of the calculations have been relegated to.

\section{Risk measures, market impact and Regularized Portfolio Optimization}
\label{Imp-Reg}

The problem we focus on is that of determining the optimal portfolio $\vec{w} = (w_1, \dots, w_N)$ of $N$ assets that an investor should hold. We assume the investor has a fixed wealth $W$ to invest and can buy assets today at current prices $\vec{p}_{\rm now}$. Therefore, the set of feasible portfolios satisfy the budget constraint
\begin{equation}
\vec{w}\cdot\vec{p}_{\rm now}=\sum_{i=1}w_i p_{i,{\rm now}}=W.
\end{equation}
The optimal portfolio is the one that minimizes the risk of the portfolio $\vec w$ at a future time, at prices $\vec{p}_{\rm est}$. Here we assume that
\begin{eqnarray}
\vec{p}_{\rm est} = \vec{p}_{\rm now} + \vec{X} -  \eta \vec{\psi}(\vec{w}),
\end{eqnarray}
where $\vec{X}$ is a random vector of returns, $\vec{\psi}(\vec{w})$ denotes the market impact function, and $\eta$ is a proportionality factor.\footnote{Notation: we use uppercase letters for random variables. {\cor $\psi_i(\vec w)$ denotes the $i^{\rm th}$ component of the vector $\vec\psi(\vec w)$.}} The term  $\eta \vec{\psi}(\vec{w})$ accounts for the finite liquidity of the market.
One way to motivate it is to argue that, in order to derive a monetary value from the portfolio, it ought to be sold \cite{CSMK}. Then this term accounts for effects of finite liquidity, which are currently of great interest \cite{Toth11,Farmer13}. The liquidation of the portfolio $\vec w$ will generally move prices ``against'' the trading activity (i.e. $\psi_i(\vec w)\ge 0$).  
Much recent work has been devoted to deriving quantitative estimates of the market impact function $\psi_i(\vec w)$ (see e.g. \cite{Toth11,Farmer13}). Most of this literature has been concerned with the single asset setting ($N=1$), suggesting that $\psi_i(\vec w) \propto w_i^\gamma$ with $\gamma$ taking values around $1/2$. 

The value of the portfolio can then be estimated by the cash flow that can be generated when the portfolio is liquidated:
\begin{eqnarray}
V(\vec w) &=& \vec{w}  \cdot\vec{p}_{\rm est}\\
&=& \vec{w} \cdot \vec{p}_{\rm now} + \vec{w}  \cdot\vec{X} - \eta \vec{w} \cdot\vec{\psi}(\vec{w}). \label{value}
\end{eqnarray}
The optimal portfolio should be chosen in such a way as to minimize the risk attached to this cash flow \cite{CSMK}. 

We discuss in section \ref{crm} how risk is measured. For the moment, we remark that 
risk enters only in the term $\vec{w} \left( \vec{X} - \eta \cdot\vec{\psi}(\vec{w}) \right)$, because $\vec{w} \cdot \vec{p}_{\rm now}=W$ is known. Hence, without loss of generality and for the sake of simplifying mathematical expressions, we will omit this term in future discussion. Also, for the sake of simplicity, we normalize current prices to $p_{i,{\rm now}}=1/N${\cor , $\forall i$, so that the budget constraint becomes
\begin{equation}
\sum_{i=1}w_i =NW.
\end{equation}
}

\subsection{Coherent risk measures} 
\label{crm}
A risk measure is a function of the random variable $Z=\vec w\cdot\vec X$, denoted by $\rho[Z]$ \footnote{
The squared parenthesis denotes that this is an operator $\rho:~\mathcal{X}\to \mathbb{R}$ that associates a real number with any random variable $Z\in\mathcal{X}$.} 
representing a loss, i.e. it is a quantity that one would like to minimize. 
Here we focus on coherent risk measures, a broad class of quantitative measures for risk \cite{Artzner99}. 
A risk measure is coherent if it satisfies the following axioms: 
\begin{description}
  \item[Normalization] $\;\rho[0]=0$.
  \item[Monotonicity] $\;$ if $Z_1\le Z_2$ a.s. then $\rho[Z_1]\ge\rho[Z_2]$
  \item[Sub-additivity] $\;\rho[Z_1+Z_2]\le \rho[Z_1]+\rho[Z_2]$ 
  \item[Positive homogeneity] $\;$ if $a\ge 0$ then $\rho[aZ]=a\rho[Z]$
  \item[Translation invariance] $\;$ if $a$ is a real constant then $\rho[Z+a]=\rho[Z]-a$
\end{description}
The failure of the widely used risk measure Value at Risk to satisfy subadditivity was the prime motivation for the introduction of these axioms \cite{Artzner99}. The translation invariance property implies that adding a certain amount $a$ (of cash) to a portfolio reduces risk by $a$.

\subsection{Market impact and regularization} 

Because of the translation invariance property, we have for any coherent risk measure, evaluated on the {\em value}, $V$, of a portfolio (see Eq. \ref{value})
\begin{eqnarray}
\rho[V] = \rho[\vec{w}\cdot\vec{X}-\eta \vec{w}\cdot\vec{\psi}(\vec{w})]=
\rho[\vec{w}\cdot\vec{X}] +\eta \vec{w}\cdot\vec{\psi}(\vec{w}).
\end{eqnarray}

The first term corresponds to the risk evaluated on the portfolio without market impact considerations. The second term, $\eta \vec{w}\cdot\vec{\psi}(\vec{w})$, is a monotonically increasing function of the weights, therefore it penalizes large positions, which means that {\em it acts as a regularizer}. It is very natural that anticipation of future liquidation should limit the sizes of one's positions. 

In summary, if the investor minimizes the risk of the cash flow that could be generated by the liquidation of the portfolio, then considering the impact of the investor's own actions on the market naturally leads to Regularized Portfolio Optimization (RPO) {\em for all coherent risk measures}. 

The correspondence between market impact functions and regularizers is as follows:
\begin{description}
  \item[The bid-ask spread impact] $\vec{\psi}(\vec{w})= {\rm sign}(\vec{w})$ leads to the use of the $L_1$-norm, which we can write as $\vec{w} \cdot {\rm sign}(\vec{w}) = \|w\|_1$. This applies to single trades, as well as to sequential trades executed in a very short time.
  \item[Square root impact] Large positions are usually liquidated by chopping the execution into many small child orders that are executed sequentially. Such {\em meta-orders} typically have an impact obeying a square root law $\psi_i(\vec{w})=\sqrt{|w_i|}$ \cite{Toth11}. This leads to a $p=3/2$ regularizing term: $\eta \vec{w} \cdot \vec\psi (\vec{w})= \eta \sum_i |w_i|^{3/2}$.
  \item[Linear impact] On longer time scales, when the sales are repeated over a long time, we expect the impact to be linear: $\vec{\psi}(\vec{w})= \vec{w}$. This leads to the regularizing term $\eta \vec{w} \cdot \vec{w} \;=\|w\|_2$, i.e. implies the use of the $L_2$ norm. 
\item[In general] $\psi_i(\vec{w})=|w_i|^{p-1}$ is equivalent to a regularizing term $\eta \sum_i|w_i|^{p}$, and thus implies the use of the $L_{p}$-norm  in RPO:
\end{description}
\begin{eqnarray*}
\hbox{Market impact:~~} \psi_i(\vec{w})=|w_i|^{p-1}~~\Leftrightarrow~~ \hbox{Regularizer}: ~~L_{p}(\vec{w})
\end{eqnarray*}

\section{{\cor Stability of $L_p$ regularized portfolio optimization}}
\label{L1-inst}
{\cor In this section we analyze the stability of $L_p$ norm based regularized portfolio optimization. We focus particularly on $p=1$, as it plays a special role. In particular, we show that, for all coherent risk measures, if $p > 1$, then regularization based on the $L_p$ norm removes the divergence that plagues portfolio optimization in the {\cor undersampled} regime, but this is not the case for $p \le 1$.}

Let us recall that the phenomenon of interest is the instability of the empirical estimate of the risk measure, when the sample size, $T$, is of the same order as the number of assets, $N$, in the portfolio, i.e. the dimensionality of the return vector.
{\cor Assume that a sample, $\mathcal{X}$, consists of $T$ observations, $\vec{x}_t$, $t=1,\ldots,T$, drawn i.i.d. from a true (but unknown) underlying probability distribution, $p$, and that there is a space of samples, ${\bf X}$, which come from the same underlying distribution.} The empirical estimate of a risk measure can then be obtained by using the empirical
distribution
\begin{equation}
\label{ }
\hat p(\vec x)=\sum_{t=1}^T\frac{\delta(\vec x-\vec x_t)}{T}~.
\end{equation}
The empirical risk, $\hat \rho$, computed on this distribution, depends on the specific sample $\mathcal{X}=\{\vec x_t\}_{t=1}^T$.
For some samples, the function $\hat\rho [\vec w]$ is unbounded from below, which means that the portfolio optimization problem does not have a finite solution. This results in an instability of the risk measure.

To develop an intuition, we first consider the simple Maximal Loss (ML) risk measure, regularized by the $L_1$ norm, for $N=2$ stocks and $T=2$ observations (Section \ref{2assets}). 

We show in Section \ref{corhR} that any coherent risk measure can become unstable to estimation error for $L_1$ norm based regularization.
The proof generalizes that given in \cite{kondor2008} for unregularized portfolio optimization. The proof relies on the probability of the existence of dominating portfolios. Depending on the ratio $N/T$, this probability can be either very close to zero -- making the instability a rare event that can be neglected -- or very close to one. Sharp results for the threshold value of $N/T$ that separates the two cases can be given in the limit $N, T\to\infty$ with fixed ratio, using methods from statistical mechanics (see Section \ref{L1ES}).  

\subsection{$L_1$ regularized ML -- a toy example}\label{2assets}

To develop an intuition about the regularized portfolio optimization problem with an $L_1$ regularizer, we first consider a simple risk measure called Maximal Loss (ML), defined as \cite{Young1998}
\begin{equation}
\label{ML}
{\rm ML}(\w)=\min_{w_i}\max_{t=1,\ldots,T} \left[-\sum_i w_i x_{i,t} \right].
\end{equation}

Maximal Loss is the best combination of the worst losses. As such it is a min-max type risk measure, a special case of Expected Shortfall (ES). Expected Shortfall is the average loss beyond a high quantile. The threshold $\alpha$ beyond which the average loss ES is calculated has typical values in the range (0.95 to 1.00). Maximal Loss corresponds to $\alpha=1-1/T$. 

The regularized portfolio optimization problem then becomes
\begin{equation}
\min_{w_i}  \left\lbrace \max_{t=1,\ldots,T} \left[-\sum_i w_i x_{i,t} \right] + \eta \|\vec{w}\|_p \right\rbrace,
\end{equation}
where $\|\vec{w}\|_p$ is the $L_p$ norm. This can be re-written to incorporate the regularizer directly into the loss function, since the term $ \|\vec{w}\|_p$ does not depend on the individual returns, measured at times $t=1, \dots, T$. We can write the {\em Regularized Maximal Loss} as
\begin{equation}
{\rm RML}(\w)=\min_{w_i}\max_{t=1,\ldots,T} \left[-\sum_i w_i x_{i,t}+\eta \|\vec{w}\|_p \right]
\end{equation}

Now, we investigate stability resulting from use of the  $L_1$ norm, $\|\vec{w}\|_1 = \sum_i | w_i | $. To that end, 
let us consider the simple case of two assets ($i=1,2$) and two data points taken at times $t=1$ and $t=2$. 
The loss associated with each time can be expressed as
\begin{eqnarray}
l_t = -wx_{1t}-(1-w)x_{2t}+\eta|w|+\eta|1-w|
\end{eqnarray}
where we have taken the budget constraint into account by setting $w_1=w$ and $w_2=1-w$
Now we distinguish three different cases:
\begin{itemize}
\item if $w>1$, then $l_t=w(x_{2t}-x_{1t}+2\eta)-x_{2t}-\eta$
\item if $0<w<1$, then $l_t=w(x_{2t}-x_{1t})+\eta-x_{2t}$
\item if $w<0$, then $l_t=w(x_{2t}-x_{1t}-2\eta)+\eta-x_{2t}$
\end{itemize}
The instability may arise in the following situations:
\begin{itemize}
\item if $w<0$, $x_{21}-x_{11}>2\eta$ and $x_{22}-x_{12}>2\eta$ (the two straights lines, $l_1$ and $l_2$, both have positive slope).
\item if $w>1$,  $x_{21}-x_{11}<-2\eta$ and $x_{22}-x_{12}<-2\eta$  (the two straights lines, $l_1$ and $l_2$, both have negative slope).
\end{itemize}

\begin{figure}[H]
\begin{center}$
\begin{array}{cc}
\includegraphics[width=4.5cm]{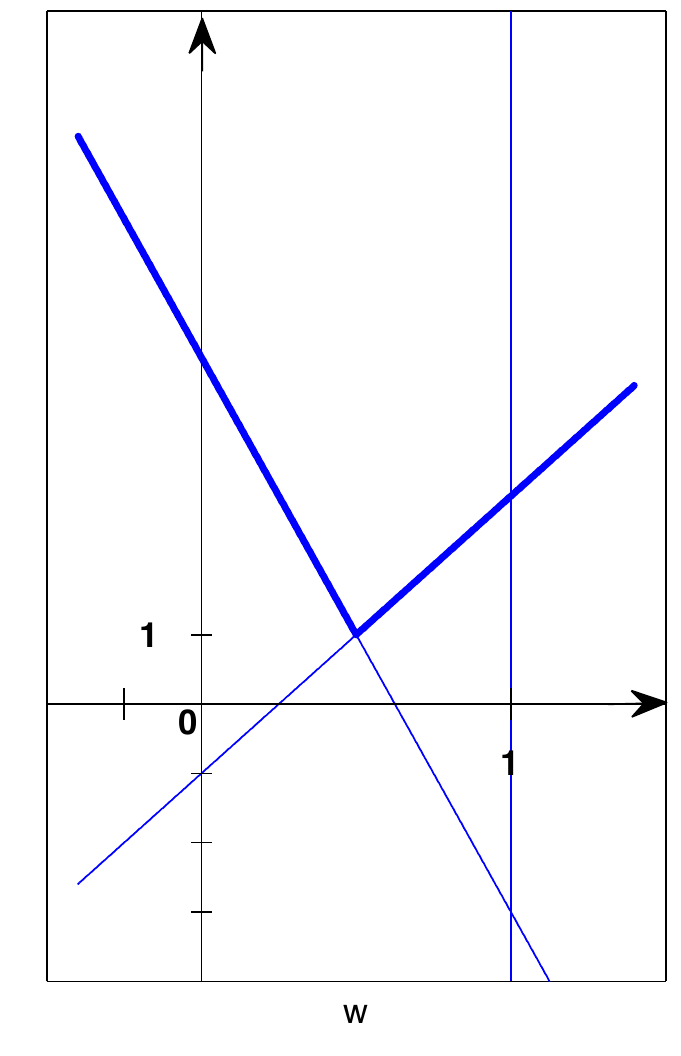}
\includegraphics[width=4.5cm]{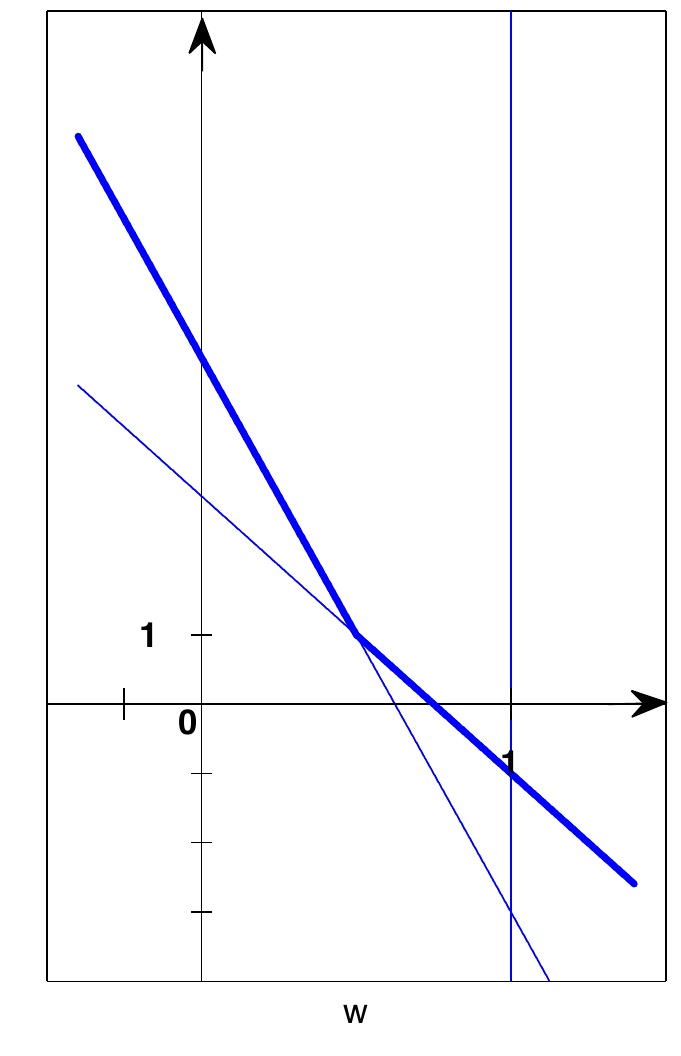}\\
\includegraphics[width=4.5cm]{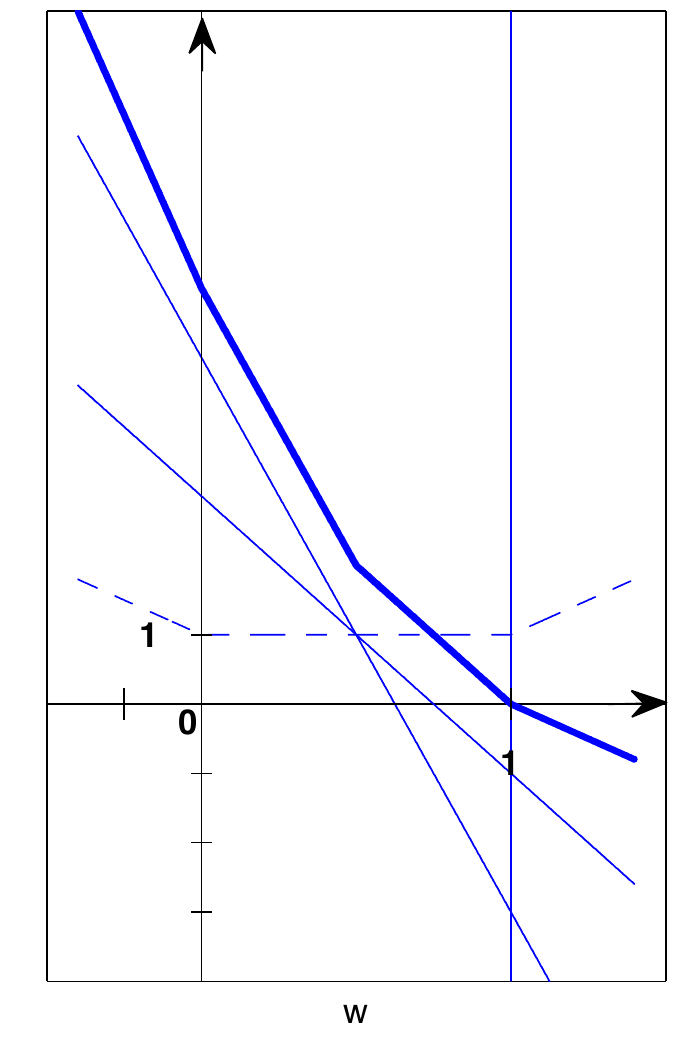}
\includegraphics[width=4.5cm]{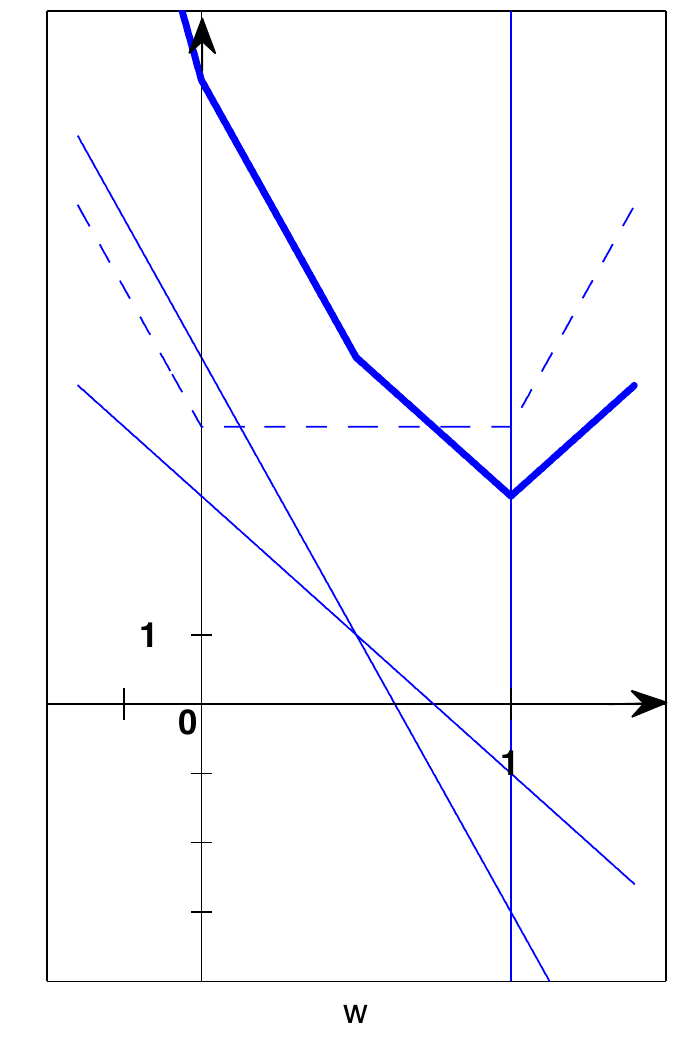}\\
\end{array}$
\end{center}
\caption{\footnotesize{ Maximal Loss with and without $L_1$ regularizer for $N=T=2$. Light solid lines represent losses at time $t = 1$ and $t = 2$. Bold lines represent {\cor the
cost function of the respective optimization problem}. The vertical solid line is drawn {\cor at}
$w = 1$. Dashed lines in the bottom panels represent the contribution of the regularizer to the cost
function.
{\bf Top left}.: Maximal Loss with $N=T=2$ when neither of the assets dominate, so a finite solution exists (the loss function has a finite minimum). 
{\bf Top right}.: Asset 1 is dominating asset 2 here, the loss function does not have a finite minimum, and the solution runs off to infinity.
{\bf Bottom left}: Same as top right, with an $L_1$ regularizer added with a coefficient eta=1. This regularizer is too weak to prevent a run-away solution.
{\bf Bottom right}.: Same as top right, but now with a strong enough regularizer (eta=4) that keeps the solution finite. Note that the regularizer sets the weight of the dominating asset to one, and that of the dominated one to zero.}}
\label{toy}
\end{figure}

This means that in the absence of regularization the Maximal Loss is unbounded from below whenever one of the assets dominates the other (i.e. its return is higher at both times). Since both the loss function and the regularizer are piecewise linear functions, there is a threshold value $\eta_c$ for any given realization of returns such that if $\eta>\eta_c$ the instability is removed (see Figure \ref{toy}). However, the threshold is in general different for different instances of the problem, therefore if we consider the whole ensemble of possible instances, the regularizer with a fixed, finite value of $\eta$ will only prevent the instability with a certain probability.

Let us calculate this probability by considering the ensemble of all possible realizations of returns. Assuming that $x_{i,t}$ are independent normal distributed random variables, the probability that the instability is not removed, is given by
 \be
 P(\eta)=\left(\int_{2\eta}^{\infty} \frac{e^{-z^2/4}}{\sqrt{4\pi}} dz \right)^2 +\left(\int_{-\infty}^{-2\eta} \frac{e^{-z^2/4}}{\sqrt{4\pi}} dz \right)^2=\frac{1}{2}{\rm erfc}^2(\eta).
 \ee
Here the integration variable $z$ is distributed as the difference of two independent Gaussian variables.

\begin{center}
\begin{figure}[h]
\begin{center}
\includegraphics[width=10cm]{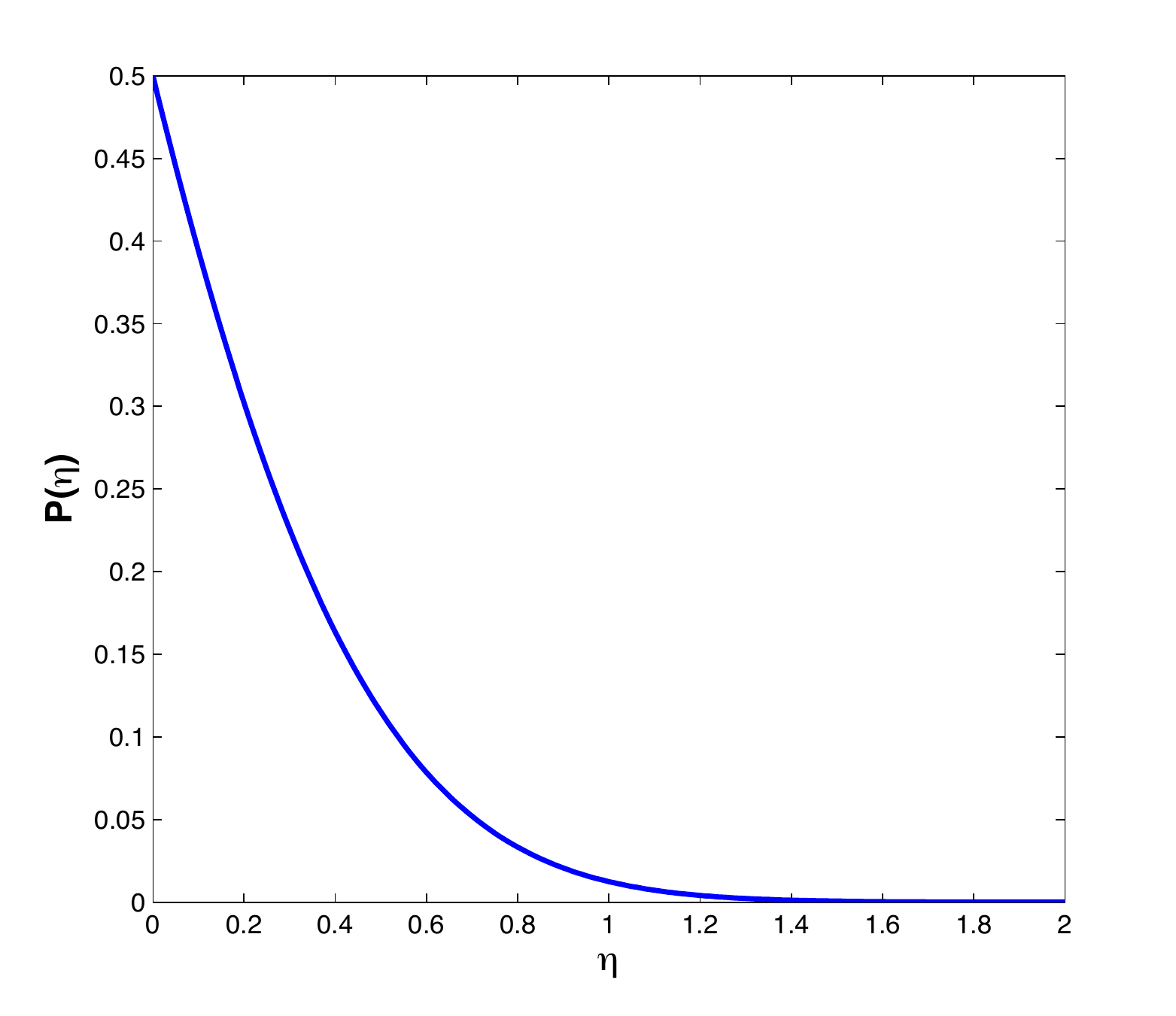}
\caption{\footnotesize {\textit{Probability that the $L_1$-regularized {\cor ML with two assets} is unstable as a function of $\eta$.}}}
\label{P_eta_2}
\end{center}
\end{figure}
\end{center}

The quantity $P(\eta)$ is plotted in Figure \ref{P_eta_2}.

This probability is positive for any finite $\eta$, which means that a "soft" implementation of the $L_1$ regularizer cannot prevent the instability. The instability disappears only for $\eta\to\infty$, when the regularizer imposes an infinite penalty on solutions with the weights outside the interval [0,1]. This "hard" constraint is equivalent to a ban on short selling.

Whenever the original problem is unstable and the $L_1$ regularizer removes the instability, it does so by setting the weight of the dominant asset to 1, and that of the dominated asset to 0. The well-known property of $L_1$ -- that it produces a sparse solution -- means in the toy example considered here that it eliminates one of the assets. 
 
The soft implementation of $L_1$ fails to eliminate the instability, because the original loss function and the regularizer are piecewise linear functions. In contrast, a nonlinear regularizer, such as the  $L_2$ norm, removes the instability, as can be seen by considering the losses at times $t=1,2$:
\begin{eqnarray} 
l_t&=&-wx_{1t}-(1-w)x_{2t}+\frac{{\eta}}{2}w^2+\frac{{\eta}}{2}(1-w)^2\\
&=&{\eta} w^2 + (x_{2 t}-x_{1 t}-{\eta})w-x_{2 t}+\frac{{\eta}}{2}
\end{eqnarray}

The losses are now convex quadratic functions, therefore a finite and unique minimum for the Maximal Loss exists for all nonzero values of ${\eta}>0$. The same holds true for any $L_p$ norm with $p>1$, whereas a norm with $p<1$ can not remove the instability.

\subsection{{\cor Instability of $L_1$ regularized coherent risk measures}}
\label{corhR}

We generalize the proof of the instability of coherent risk measures \cite{kondor2008}, which relies on the notion of dominant portfolios. The first step is to show that if such a dominant portfolio exists, then coherent risk measures are unbounded from below and the optimization problem is ill-defined. The second step is to show that the probability that dominant portfolios exist is bounded away from zero.

A vector $\vec u$ is said to be a $\mu$-dominant portfolio if {\em i)} $\vec u \cdot \vec 1=0$ and {\em ii)} $\vec u \cdot \vec x_t\ge\mu |\vec u|$ for all $t=1,\ldots,T$, where $|\vec u|=\sum_i|u_i|$ is the $L_1$ norm.

\vskip 0.5cm

{\bf Proposition:} If a $\mu$-dominant portfolio exists, then no finite $L_1$-regularized portfolio can exist for $\eta<\mu$.
\vskip 0.5cm

{\em Proof:}  The objective function that we wish to minimize is 
\begin{equation}
\label{ }
\hat\rho[\vec w\cdot\vec X]+\eta|\vec w|.
\end{equation}

For any normalized portfolio $\vec w$ (i.e. $\vec w\cdot\vec 1=1$) consider the portfolio $\vec w+a\vec u$, where $\vec u$ is a $\mu$-dominant portfolio. This is a normalized portfolio, because of {\em i)} in the definition of $\vec u$. Now
\begin{eqnarray}
\hat\rho[(\vec w+a\vec u)\cdot \vec{X}] & \le & \hat\rho[\vec w\cdot X]+\hat\rho[a\vec u\cdot X] ~~~\qquad\hbox{sub-additivity}\nonumber\\
 & = &  \hat\rho[\vec w\cdot X]+a\hat\rho[\vec u\cdot X] ~~~\qquad\hbox{positive homogeneity}\nonumber\\
 & \le &  \hat\rho[\vec w\cdot X]+a\hat\rho(\mu|\vec u|) ~~~~\qquad\hbox{monotonicity ~} (a>0) \label{proof.3}\\
 & = &  \hat\rho[\vec w\cdot X]-a\mu|\vec u| ~~\qquad\qquad\hbox{translation inv. $+$ normalization.}\nonumber
\end{eqnarray}
To get to eq. (\ref{proof.3}), we used monotonicity with $Z_1=\mu$ (constant) and $Z_2=\vec u\cdot\vec X$. 

For positive constant $a$, $|\vec w+a\vec u|\le |\vec w|+a|\vec u|$ which means that 
\begin{equation}
\hat\rho[(\vec w+a\vec u)\cdot\vec X]+\eta|\vec w+a\vec u|  \le  \hat\rho[\vec w\cdot X]+\eta|\vec w|-a(\mu-\eta)|\vec u| \\
\end{equation}
For $\mu>\eta$ we can achieve portfolios with negative risk of arbitrarily large absolute value by letting $a\to\infty$: the solution of the optimization problem runs off to infinity. $\Box$

Therefore,  $L_1$ regularization is not a good choice if the probability that a $\mu$-dominant portfolio exists is non-zero. The probability that a $\mu$-dominant portfolio exists depends on the distribution of returns, and it is therefore hard to make any general statements without making assumptions about this distribution. Yet, if the support of the distribution of returns extends to the whole real axis, then the probability that a $\mu$-dominant portfolio exists is non-zero. In the next section we assume Gaussian (random) portfolios, providing some intuition about this probability.

Notice that the proof breaks down when the $L_1$ norm is replaced by the $L_p$ norm with $p>1$. Then the objective function
\begin{equation}
\label{ }
\hat\rho[\vec w\cdot\vec X]+\eta\sum_{i=1}^N|w_i|^p\sim \eta\sum_{i=1}^N|w_i|^p \left(1+O(w^{-(p-1)})\right)
\end{equation}
is dominated by the regularization term when the weights $w_i$ diverge. 

Conversely, for $p<1$ it is the regularization term that becomes negligible compared to the risk measure. Hence regularization with $p<1$ is not able to cure the instability, as soon as a $\mu$-dominant portfolio arises, for any $\mu\ge 0$.

\section{Regularized Expected Shortfall for large random portfolios.}
\label{L1ES}

In this section we consider the problem of RPO for the risk measure Expected Shortfall in the particular case where both the number of assets $N$ and the number of observations $T$ are large ($N,T\to\infty$), but their ratio is a finite constant. In order to derive quantitative results, we consider, for the sake of simplicity, the case of $N$ i.i.d standard normal returns. We can extend the approach of \cite{CKM07} to include regularization (see also \cite{CSMK}). This provides us with typical results, i.e. results that occur with probability arbitrarily close to one in the limit $N,T\to\infty$.

For i.i.d. random returns, the true optimal solution is the one where all the weights are the same, $w_i = 1/N, \; \forall i$. Individual solutions based on specific samples will likely deviate from the true optimum, especially when the sample size, $T$, is not large enough. The optima for individual samples may deviate from symmetry, but averaging over the samples will necessarily restore it. Likewise, the $L_1$ regularizer that tends to produce sparse solutions will not be able to make a distinction between the equivalent assets, therefore it will set some weights to zero at random for each sample. Our goal here is merely to demonstrate the effect of $L_1$ on the instability of a portfolio optimized under ES. The real effect of $L_1$ should be studied using a heterogeneous (non-i.i.d.) portfolio, a problem we leave for future research.

\begin{center}
\begin{figure}[h]
\begin{center}
\includegraphics[width=10cm]{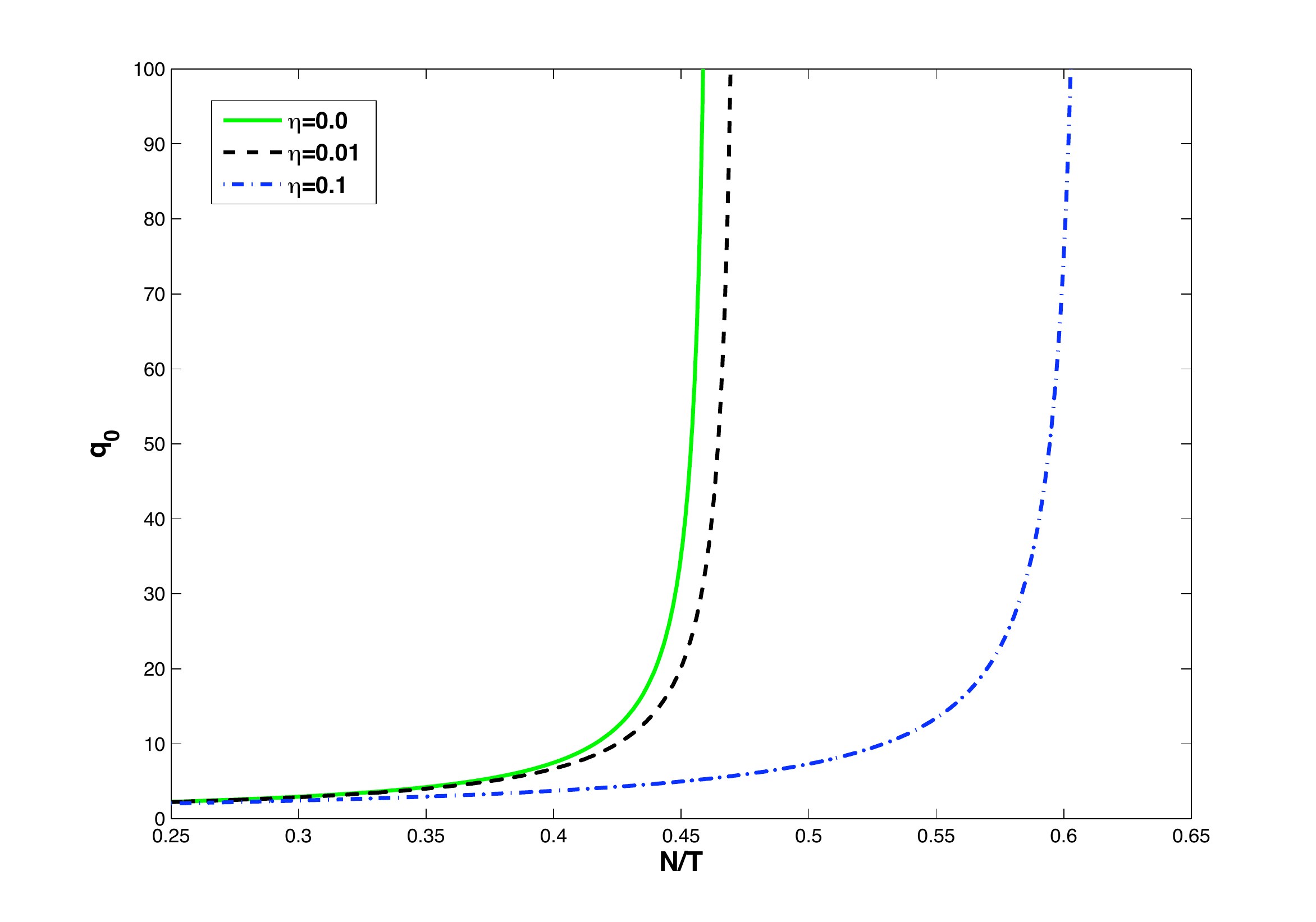}
\caption{\footnotesize {\textit{$q_0$ as a function of {\cor$N/T$} for different values of $\eta$ for the $L_1$ regularized ES. The $L_1$ regularizer shifts the feasible-{\cor i}nfeasible transition towards {\cor larger} values of $N/T$.} } }
\label{L1_est_ES}
\end{center}
\end{figure}
\end{center}

The optimization of historical Expected Shortfall can be formulated as a linear programming problem \cite{Rockafellar}. Expected Shortfall is a piecewise linear loss function, just like the Maximal Loss. 
The optimization problem, including the $L_p$ regularizer, reads 
\begin{eqnarray}
&&\min_{\w, {\vec u}, \epsilon}  \left[ (1-\beta) T\epsilon+\sum_{t=1}^T u_{t}+{\eta } |\w|^p\right],
\label{newPO} \\
&{\rm s.t.}\;\;\; & \vec{w}\cdot\vec{x}_t + \epsilon + u_t \geq 0; \;\;\; u_t \geq 0; \;\;\; \forall t, \label{constraints} \\
&& \sum_i w_i = WN, \label{b}
\end{eqnarray}
where $W$ is a normalization parameter that is usually set to one. However, since market-impact depends on the size of the position that is liquidated, we explicitly keep track of $W$ in what follows. In addition, the sum of the weights was chosen to be of the order of $N$, in order for the optimal weights to be of order one.

Assuming Gaussian returns, we can solve the above optimization problem analytically. Generalizing \cite{CKM07, CSMK} by including the regularizer, we follow lines familiar from the statistical physics of disordered systems: the loss function is regarded as the Hamiltonian (energy functional) of a fictitious statistical physics system, a fictitious temperature is introduced and the free energy of this system is calculated in the limit $N,T\to\infty$ with $T/N=\tau$ fixed. Averaging over the different realizations of the returns corresponds to what is called quenched averaging in statistical physics. We report here only the final result, and refer the interested reader to the Appendix, where details of the calculation are given. After some effort, one ends up with a free energy  functional, depending on six variables, the so-called order parameters:
{\cob
\bea
\label{free_energy}
F({\lambda},{\epsilon},{q}_0,\Delta, {\hat{q}}_0,\hat{\Delta})&=&
{\lambda} W +\tau (1-\beta)\epsilon -\Delta{\hat{q}}_0-\hat{\Delta}{q}_0\\
\nonumber &+& \langle {\rm min}_w \left[V(w,z)\right]\rangle_z
 +\frac{\tau\Delta}{2\sqrt{\pi}}\int_{-\infty}^{\infty}ds e^{-s^2}
g\left({\frac{\epsilon}{\Delta}}+s \sqrt{\frac{2{q}_0}{\Delta^2}} \right),
\eea
where 
\begin{equation}
\label{pot}
V(w,z)=\hat{\Delta} w^2 + \eta |w|^{p}-{\lambda} w -z w\sqrt{-2{\hat{q}}_0}.
\end{equation}
}
The function $g(x)$ is given in the Appendix. In Eq. (\ref{free_energy})  $\langle\cdot\rangle_z$ represents an average over the normal variable $z$.
The ``true" free energy, i.e. the minimal risk per asset,  is the stationary point of the above expression with respect to variations of the order parameters ${\lambda},{\epsilon},{q}_0,\Delta, {\hat{q}}_0$ and $\hat{\Delta}$.
The original optimization problem involving $N+T+1$ variables has been reduced, in the limit of large $N$ and $T$, to the simpler problem of finding the minimum of the functional $F({\lambda},{\epsilon},{q}_0,\Delta, {\hat{q}}_0,\hat{\Delta})$ that only depends on six variables.

\subsection{$L_1$ regularization}

The origin of the instability for the Expected Shortfall under $L_1$ is the same as with Maximal Loss: Expected Shortfall diverges (becomes unbounded from below) when there is a dominating portfolio \cite{kondor2008}. The divergence of ES is linear, therefore a piecewise linear regularizer like $L_1$ will not be able to eliminate the instability completely, it will do so only with a certain probability, depending on the sample. Thus we expect that after averaging over the ensemble of random returns the instability will persist. Yet, we know \cite{CKM07} that the probability to observe the instability is essentially zero for $N,T\to \infty$ when $\tau=T/N$ is larger than a critical threshold $\tau_c$ whereas it is arbitrarily close to one if $\tau<\tau_c$.
We now illustrate how the instability arises by analyzing the extrema of Eq. (\ref{free_energy}), and show that $L_1$ regularization merely shifts the instability to a smaller threshold.

{The fact that the solution for $N,T\to\infty$ contains a nested optimization problem over $w$ in the first term on the second line of Eq. (\ref{free_energy}) is not accidental. Indeed, this arises from the optimization over the weights $w_i$ of the original problem. Therefore, the minimization of $V(w,z)$ can be thought of as a ``representative weight'' problem, where the random variable $z$ over which the expectation is taken encodes the effect of the randomness in the sample. 

The solution $w^*(z)$ of the representative weight problem
\begin{equation}\label{wstar}
    w^*(z)= \left\{ \begin{array}{cc} 
    \frac{z\sqrt{-2\hat{q}_0}+\lambda-\eta}{2\hat{\Delta}} ,& {\rm if~~}    z>\frac{\eta-\lambda}{\sqrt{-2\hat{q}_0}}\\
    0 ,& {\rm if~~}    \frac{-\eta-\lambda}{\sqrt{-2\hat{q}_0}}<z<\frac{\eta-\lambda}{\sqrt{-2\hat{q}_0}}\\
 \frac{z\sqrt{-2\hat{q}_0}+\lambda+\eta}{2\hat{\Delta}} ,& {\rm if~~}    z<\frac{-\eta-\lambda}{\sqrt{-2\hat{q}_0}} ,    \end{array} \right.
\end{equation}
has the typical shape induced by $L_1$ regularization. The Gaussian distribution of the $z$ then induces the weights to be distributed according to $p(w)=\langle \delta(w-w^*(z))\rangle_{z}$.

The first-order conditions for the free energy read
{\cob
\be
W=\avg{w^*}_{z}\label{spBudget}
\ee
\be
(1-\beta)+\frac{1}{2\sqrt{\pi}}\int_{-\infty}^\infty ds e^{-s^2}g'\left(\frac{\epsilon}{\Delta}+s\sqrt{\frac{2{q}_0}{\Delta^2}}\right)=0
\ee
\be
\hat{\Delta} - \frac{\tau}{2\sqrt{2\pi{q}_0}}\int_{-\infty}^\infty ds e^{-s^2}s g'\left(\frac{\epsilon}{\Delta}+s\sqrt{\frac{2{q}_0}{\Delta^2}}\right)=0
\ee

\be
-{\hat{q}}_0-2\frac{\hat{\Delta}{q}_0}{\Delta} + \frac{\tau}{2\sqrt{\pi}}\int_{-\infty}^\infty ds e^{-s^2}g\left(\frac{\epsilon}{\Delta}+s\sqrt{\frac{2{q}_0}{\Delta^2}}\right)+\tau(1-\beta)\frac{{\epsilon}}{\Delta}=0
\ee

\be
\Delta=\frac{1}{\sqrt{-2{\hat{q}}_0}}\avg{w^* z}_{z}\label{spDelta}
\ee

\be
{q}_0=\avg{{w^*}^2}_{z}.\label{spQ}
\ee
}

Notice in particular, that the first of these equations translates the budget constraint of the original problem into the properties of the representative weight problem above. The last equation, instead, relates the parameter ${q}_0\propto \frac{1}{N}\sum_i w_i^2$ to the fluctuation of the weights. This corresponds to the 
distance of the solution from the optimal one (given by the constant portfolio). 
In the present setting, this is also related to the estimation error, since different samples will deviate from the optimal solution in different ways. 

The instability of portfolio optimization manifests itself in the divergence of the parameter ${q}_0$. This, in turn, implies that the distribution of weights $w^*(z)$ becomes degenerate, i.e. infinitely broad. Hence, some intuition about the origin of the instability can be gained from looking at the optimization problem for the representative weight. From the form of $V(w,z)$ we see that the optimization problem for $w$ always has a finite solution as long as the order parameter $\hat{\Delta}>0$. However, when $\hat\Delta=0$, $V(w,z)$ can be unbounded from below. In this case, the estimated expected shortfall runs to minus infinity, because the term $\langle {\rm min}_w \left[V(w,z)\right]\rangle_z$ in the free energy diverges. Note that, because of the budget constraint implemented in equation \eqref{spBudget}, the optimal portfolio will in this case dictate to go infinitely long in some of the assets and infinitely short in some of the others.
In summary, the condition $\hat\Delta =0$ is what determines the critical point. 

We numerically solved the system of equations {\cob \eqref{spBudget}--\eqref{spQ}}, and we plot in Figure \ref{L1_est_ES} the quantity $q_0$ as a function of the ratio $\tau^{\cob-1} = N/T$ for different values of the regularizer's amplitude $\eta$. 
The study of the stationarity conditions shows that, similarly to what we saw in the toy example, the $L_1$ regularization produces a shift towards higher values of $N/T$ of the  feasible-infeasible transition, but does not eliminate it completely. The shift in the critical point where the estimation error blows up is clearly seen in Figure \ref{L1_est_ES}.
}

In the toy example discussed in section \ref{2assets} we furthermore saw that the hard implementation of the $L_1$ constraint, obtained in the limit $\eta\to\infty$, was equivalent to a ban on short selling. This is true also here. Indeed, if $\eta$ goes to infinity, the regularizer imposes infinite penalty on any solution not on the $L_1$ ball (defined by the relation $\sum_i w_i\le NW$). 
On the other hand, the solution also has to be on the plane corresponding to the budget constraint. This means the solution must be on the (+,+,+...+) 
face of the $L_1$ ball, where all the weights are positive. Conversely, if all the weights are positive and they also satisfy the budget constraint, then the solution is necessarily on the (+,+,+...+) face of the $L_1$ ball. 

That imposing a ban on short positions helps taming the large sample fluctuations was noticed by \cite{Jagannathan2003} though the authors did not make the connection to regularization. It is clear that any constraint that reduces the domain over which the optimum is sought to a finite volume has a similar effect: there cannot be infinite fluctuations in a finite volume. 

\subsection{$L_p$-norm}
\label{LpES}

There is no qualitative difference between the effects of the soft (finite $\eta$) or hard ($\eta\to\infty$) implementation of a non-linear regularizer that grows faster than linear, such as the $L_p$-norm with $p>1$; the regularizer will eliminate the instability either way. 

We can see this by noting that if $p>1$, then the function $V(w,z)$ always has a finite minimum for any value of $z$, thus preventing the divergence of the estimation error. 
The {\cob solution of the representative weight problem, $w^*$} can be obtained by minimization of the function $V(w,z)$ (Eq. \ref{pot}).
Let $p=1+\frac{1}{n}$. To compute $w^*$ we must solve
\be
\label{dV}
{\cob 2}\hat{\Delta} w+ \eta{\cob \left(1+\frac{1}{n}\right)} {\cob |}w{\cob ^*|}^{1/n}-{\lambda}-z\sqrt{-2{\hat{q_0}}}=0.
\ee 
Taking the limit $\hat{\Delta}\to 0$, if $\eta>0$ the equation reduces to
\be
\eta{\cob \left(1+\frac{1}{n}\right)} {\cob |}w{\cob ^* |}^{1/n}-{\lambda}-z\sqrt{-2{\hat{q_0}}}=0,
\ee 
which always has a finite solution for finite $n$.
More generally, we can say that any regularizer that grows faster than linear will prevent the solutions from running away to infinity.

This is true in particular for the for the case $p=2$ \cite{CSMK}, and for the case $p = 3/2$, which corresponds to the square root market impact that characterizes the liquidation of large positions under normal market conditions \cite{Moro09,Toth11,Farmer13,Caccioli12}. 

\subsection{Linear combination of the $L_1$ and $L_2$ regularizers.}

In high-dimensional problems it is often expedient to use a combination of the $L_1$ and $L_2$ regularizers \cite{buh11}, sometimes called the elastic net \cite{Zou05}. This setting leads to the following optimization problem 
\begin{eqnarray}
&&\min_{\w, {\vec u}, \epsilon}  \left[ (1-\beta) T\epsilon+\sum_{t=1}^T u_{t}+{\eta_1} |\w|+{\eta_2} |\w|^2\right],
\label{newPO-2} \\
&{\rm s.t.}\;\;\; & \wxk + \epsilon + u_t \geq 0; \;\;\; u_t \geq 0; \;\;\; \forall t, \label{constraints-2} \\
&& \sum_i w_i = WN. \label{b-2}
\end{eqnarray}

\begin{figure}[h]
\begin{center}$
\begin{array}{cc}
\includegraphics[width=2.75in]{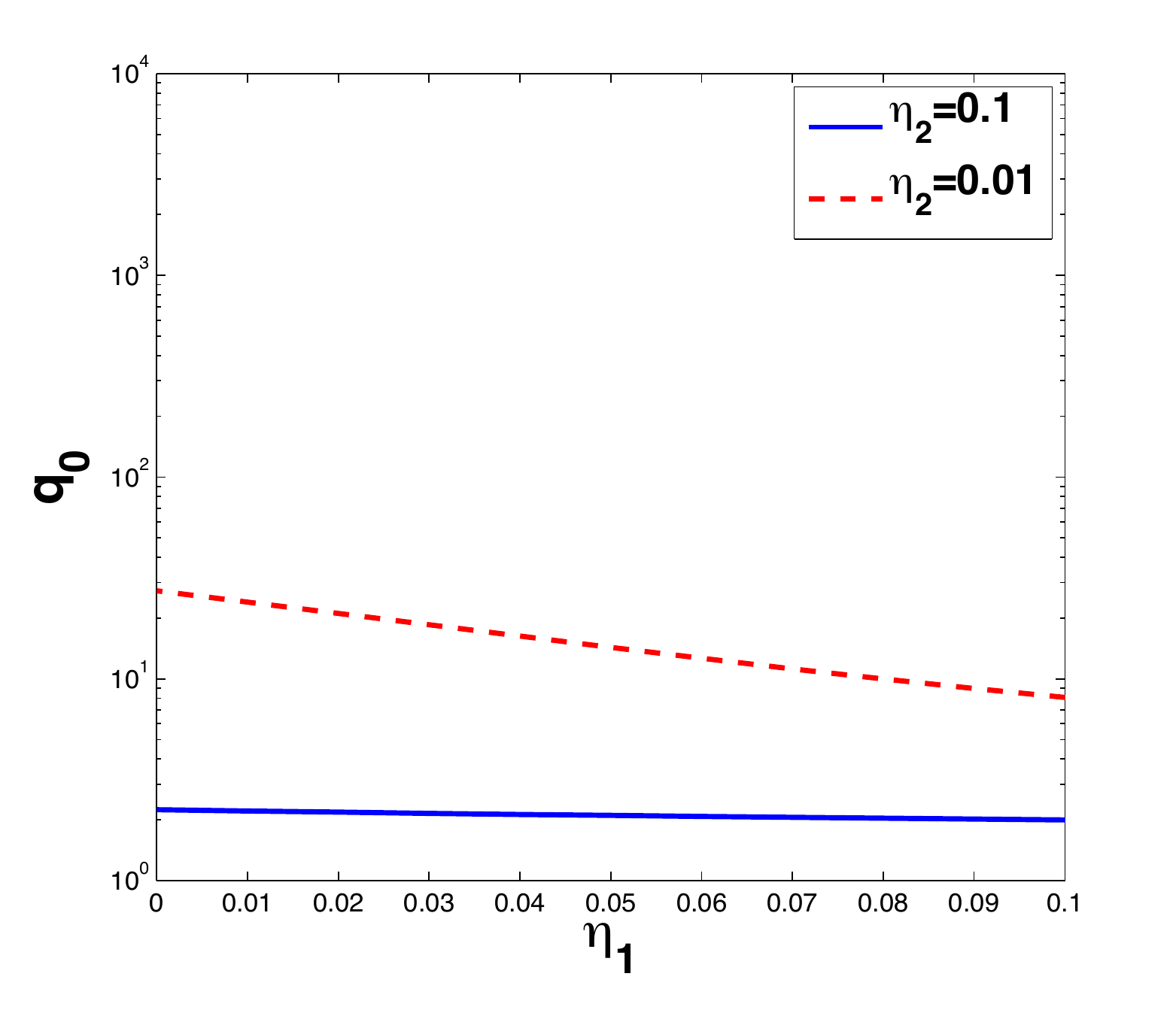}
\includegraphics[width=2.75in]{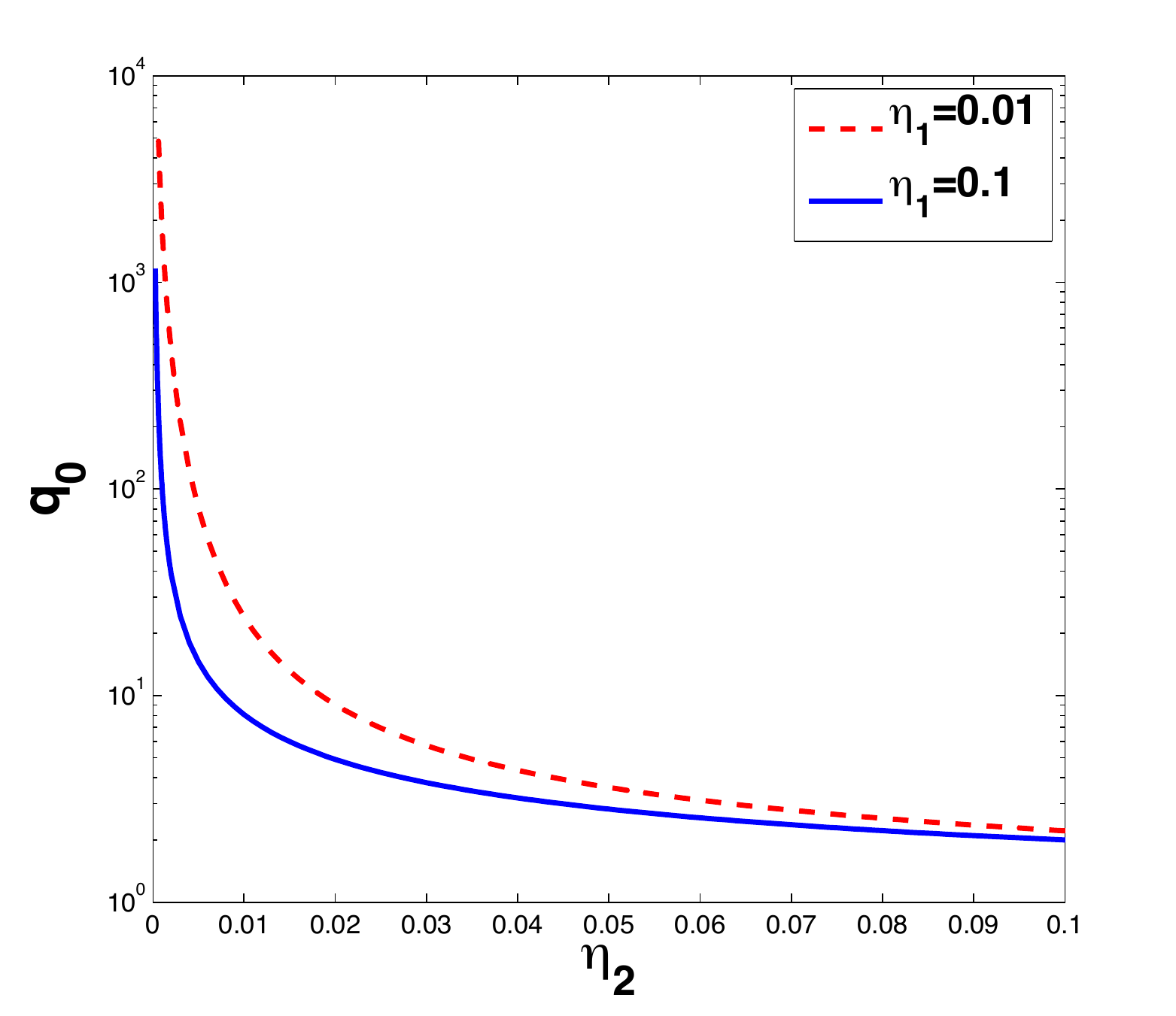}
\end{array}$
\end{center}
\caption{\footnotesize{{\bf Left panel:} $q_0$ as a function of  $\eta_1$ for $\eta_2=0.1$ (blue solid line) and $\eta_2=0.01$ (red dashed line). {\bf Right panel:} $q_0$ as a function of  $\eta_2$ for $\eta_1=0.1$ (blue solid line) and $\eta_1=0.01$ (red dashed line). In both panels $\tau=1.5$. The quantity $q_0$ strongly depends on $\eta_2$, in particular close to $\eta_2=0$.}}
\label{Lmix}
\end{figure}

The replica method readily extends also to this case. In this situation the free energy is again of the form (\ref{free_energy}), but with
\be
V(w,z)=\hat{\Delta} w^2+\eta_1|w|+\eta_2|w|^2-{\lambda} w -z w\sqrt{-2{\hat{q}}_0} ~.
\ee
The solution of the optimization problem  for the representative weight reads:
{\cob
\begin{equation}
    w^*(z)= \left\{ \begin{array}{cc} 
    \frac{z\sqrt{-2\hat{q}_0}+\lambda-\eta_1}{2(\hat{\Delta}+\eta_2)} ,& {\rm if~~}    z>\frac{\eta_1-\lambda}{\sqrt{-2\hat{q}_0}}\\
    \\
       0 ,& {\rm if~~}    \frac{-\eta_1-\lambda}{\sqrt{-2\hat{q}_0}}<z<\frac{\eta_1-\lambda}{\sqrt{-2\hat{q}_0}}\\
\\
 \frac{z\sqrt{-2\hat{q}_0}+\lambda+\eta_1}{2(\hat{\Delta}+\eta_2)} ,& {\rm if~~}    z<-\frac{\eta_1+\lambda}{\sqrt{-2\hat{q}_0}}.   \end{array} \right.
\end{equation}
}
From these expressions we see that the divergence of $w^*$ is prevented as long as $\eta_2>0$, while a divergence may occur if $\eta_2=0$ when $\hat{\Delta}=0$.

Using this form of $w^*(z)$ to evaluate the averages over the variable $z$ that appear in equations \eqref{spBudget}, \eqref{spDelta} and \eqref{spQ}, we  numerically solve the set of first-order conditions {\cob \eqref{spBudget}--\eqref{spQ}}.

We show the behavior of $q_0$ as a function of the parameters $\eta_1$ and $\eta_2$, for $\tau=1.5$, in Figure \ref{Lmix} .
The dependence on $\eta_2$ is naturally stronger than the dependence on $\eta_1$. 

A useful scheme may be to set $\eta_2$ to a value that ensures small estimation error given the value of  $\tau$ in a particular application, and then to vary $\eta_1$ in order to achieve the desired level of sparseness. Although sparseness reduces diversification, there might be situations where limiting the dimension of a portfolio may be advantageous, for example by limiting transaction costs. 

\section{Conclusion}
We have shown that  market impact considerations will drive an investor to Regularized Portfolio Optimization (RPO) under any coherent risk measure. The impact function determines which regularizer should be used. Generalizing our previous result, that linear impact leads to the use of the $L_2$ norm regularizer \cite{CSMK}, we have shown here how the impact function selects the regularizer in general. 

We have also demonstrated that the instability of Expected Shortfall found in \cite{CKM07} will be removed by any nonlinear regularizer that grows faster than linear, but will only be shifted by the piecewise linear $L_1$ norm based regularizer that corresponds to a bid-ask spread impact. Although we wanted to mostly concentrate on Expected Shortfall in this paper, we would like to point out that due to the positive homogeneity and translational invariance axioms, $L_p$ regularizers will have the same effect on any coherent measure as they do on ES. (Moreover, as these properties are shared also by Value at Risk, this extends also to VaR.) 

We find it appealing that the statistically motivated use of regularizers is intimately linked to market impact: consideration of finite liquidity should limit excessively  large positions, which is exactly what regularizers do. 

We also find it remarkable how powerful the method of replicas proves to be in producing analytic results for the estimation error in situations which would probably be very hard to approach by more conventional methods of probability theory.

\section*{Acknowledgement}
I.K. has been supported by the European Union under grant agreement No. FP7-ICT-255987-FOC-II Project and by the Institute for New Economic Thinking under grant agreement ID: INO1200019. S.S. thanks the Abdus Salam International Center for Theoretical Physics for their hospitality and support of this collaboration. M.M. acknowledges support from the Marie Curie Training Network NETADIS (FP7 - Grant 290038).

\section*{Appendix}
In this appendix we show how, in the limit where $N$ and $T$ go to infinity, the optimization problem defined by equations \eqref{newPO}, \eqref{constraints} and \eqref{b} can be reduced to that of finding the minimum of the free energy functional given by \eqref{free_energy}.

We need to find the minimum of
$$E[\epsilon,\{u_{t}\}]=(1-\beta) T\epsilon+\sum_{t=1}^T u_{t}+\eta \|\vec{w}\|_p$$
under the constraints
$$u_{t}\ge0,\;\;\;   u_{t}+\epsilon+\sum_{i=1}^N x_{i,t} w_i\ge 0\;\;\; \forall t$$ and $$\sum_{i=1}^N w_i =WN.$$
The calculation proceeds as follows:
Given a history of returns $\{x_{i,t}\}$, we introduce the inverse temperature parameter $\gamma$ and define the canonical partition function (or generating functional) as
\be
Z_\gamma\left[\{x_{i,t}\}\right]=\int_0^\infty \prod_{i=1}^T d u_t\int_{-\infty}^\infty d\epsilon~ \theta\left(u_{t}+\epsilon+\sum_{i=1}^N x_{i,t} w_i\right) e^{-\gamma E[\epsilon,\{u_{t}\}]},
\ee
where $\theta(x)=1$ if $x>0$ and zero otherwise.
The partition function is therefore an integral over all possible configurations of variables that are compatible with the constraints of the problem, where each configuration $\epsilon,\{u_{t}\}$ is weighted by the quantity $e^{-\gamma E[\epsilon,\{u_{t}\}]}$ (the Boltzmann weight).
From the partition function, the minimum cost (per asset) can be computed in the limit of large $N$ as
\be
\lim_{N\to\infty}\lim_{\gamma\to\infty}-\frac{\log Z_{\gamma}[\{x_{i,t}\}]}{\gamma N}.
\ee
To derive the typical properties of the ensemble we have to average over all possible realizations of returns and compute
\be
\langle\log Z_{\gamma}\left[\{x_{i,t}\}\right]\rangle=\int_{-\infty}^{\infty} \prod_{i=1}^N\prod_{t=1}^T dx_{i,t} P[\{x_{i,t}\}]\log Z_{\gamma}\left[\{x_{i,t}\}\right],
\ee
where $P[\{x_{i,t}\}]$ is the probability density function of returns.
This calculation can be performed by help of the replica trick, using the identity
\be
\langle \log Z\rangle = \lim_{n\to 0}\frac{\partial \langle Z^n\rangle}{\partial n}.
\ee
For integer $n$, we can compute $Z^n$ as the partition function of a system composed of $n$ replicas of the original systems. An analytical continuation to real values of $n$ will then allow us to perform the limit $n\to0$ and obtain the sought quantity $\langle\log Z_{\gamma}\left[\{x_{i,t}\}\right]\rangle$.

The replicated partition function can be computed as\footnote{In the calculation we will not keep track of constant multiplicative factors that do not affect the final result.}
\begin{eqnarray*}
Z_{\gamma}^n &=& \int_{-\infty}^{\infty}\left(\prod_{i=1}^N\prod_{t=1}^T dx_{i,t}\right)\int_{-\infty}^{\infty} \left(\prod_{a=1}^n d\epsilon^a\right)
\int_0^{\infty}\left(\prod_{t=1}^T\prod_{a=1}^n du_{t}^a\right)\int_{-\infty}^{\infty}\left( \prod_{i=1}^N \prod_{a=1}^n d
w_i^a\right)\\
&\times&\int_{-i\infty}^{i\infty}\left(\prod_{a=1}^n d\hat\lambda^a\right)  \int_{0}^{\infty} \left(\prod_{t=1}^T\prod_{a=1}^n d
\mu_{t}^a\right) \int_{-\infty}^{\infty}\left(\prod_{t=1}^T \prod_{a=1}^n d\hat{\mu}_{t}^a \right)\prod_{t=1}^T \prod_{i=1}^N\exp\left\{-\frac{Nx_{i,t}^2}{2}\right\}\\
&\times&\exp\left\{\sum_a \hat\lambda^a
(\sum_i w_i^a-WN)\right\}
\prod_{t}\exp\left\{\sum_a i \hat{\mu}_{t}^a\left(u_{t}^a+\epsilon^a+\sum_i x_{i,t}w_i^a-\mu_{t}^a \right)\right\}\\
&\times& \exp\left\{-\gamma\sum_a(1-\beta)T \epsilon^a-\gamma\sum_{a,t} u_{t}^a-\gamma\eta\sum_i
{|w_i^a|^p}\right\},
\end{eqnarray*}
where we have assumed that 
\be
P[\{x_{i,t}\}]=\prod_{t=1}^T \prod_{i=1}^N\exp\left\{-\frac{Nx_{i,t}^2}{2}\right\},
\ee 
and we have enforced the constraints through the Lagrange multipliers $\hat\lambda^a$, $\mu_t^a$ and $\hat\mu_t^a$.
Averaging over the quenched variables $\{x_{i,t}\}$ and introducing the overlap matrix
$Q_{a,b}=\frac{1}{N}\sum_i w_i^a w_i^b$  and its conjugate $\hat Q_{a,b}$ one obtains

\begin{eqnarray*}
Z_{\gamma}^n&=& \int_{-i\infty}^{i\infty}\left(\prod_{a=1}^n\prod_{b=1}^n dQ_{a,b}d\hat Q_{a,b}\right)\int_{-\infty}^{\infty} \left(\prod_{a=1}^n d\epsilon^a\right)
\int_0^{\infty}\left(\prod_{t=1}^T\prod_{a=1}^n du_{t}^a\right)\int_{-\infty}^{\infty}\left( \prod_{i=1}^N \prod_{a=1}^n d
w_i^a\right)\\
&\times&\int_{-i\infty}^{i\infty}\left(\prod_{a=1}^n d\hat\lambda^a\right)  \int_{0}^{\infty} \left(\prod_{t=1}^T\prod_{a=1}^n d
\mu_{t}^a\right) \int_{-\infty}^{\infty}\left(\prod_{t=1}^T \prod_{a=1}^n d\hat{\mu}_{t}^a \right)\exp\left\{\sum_a
\hat\lambda^a(\sum_i w_i^a-WN)\right\}
\\
&\times& \prod_{t}\exp\left\{-\frac{1}{2}\sum_{a,b}\hat{\mu}_{t}^aQ_{a,b}\hat{\mu}_{t}^b\right\}
 \exp\left\{\sum_{a,b}\hat{Q}_{a,b}\left(N Q_{a,b}-\sum_i w_i^a w_i^b\right)\right\}\\
&\times& \exp\left\{-\gamma\sum_a(1-\beta)T \epsilon^a-\gamma\sum_{a,t}
u_{t}^a-\gamma\eta\sum_i
{|w_i^a|^p}\right\}\\
&\times&\prod_{t}\exp\left\{i\sum_a\hat{\mu}_{t}^a\left(u_{t}^a+\epsilon^a-\mu_{t}^a \right)\right\}.
\end{eqnarray*}

We can now perform the Gaussian integral over the variables $\{\hat{\mu}_{t}^a\}$:
\begin{eqnarray*}
Z_{\gamma}^n&=&  \int_{-i\infty}^{i\infty}\left(\prod_{a=1}^n\prod_{b=1}^n dQ_{a,b}d\hat Q_{a,b}\right)\int_{-\infty}^{\infty} \left(\prod_{a=1}^n d\epsilon^a\right)
\int_0^{\infty}\left(\prod_{t=1}^T\prod_{a=1}^n du_{t}^a\right)\int_{-\infty}^{\infty}\left( \prod_{i=1}^N \prod_{a=1}^n d
w_i^a\right)\\
&\times&\int_{-i\infty}^{i\infty}\left(\prod_{a=1}^n d\hat\lambda^a\right)  \int_{0}^{\infty} \left(\prod_{t=1}^T\prod_{a=1}^n d
\mu_{t}^a\right) \exp\left\{\sum_a
\hat\lambda^a(\sum_i w_i^a-WN)\right\}\\
&\times& \exp\left\{-\gamma\sum_a(1-\beta)T \epsilon^a-\gamma\sum_{a,t}
u_{t}^a-\gamma\eta \sum_i{|w_i^a|^p}\right\}\exp\left\{\sum_{a,b}\hat{Q}_{a,b}\left(N Q_{a,b}-\sum_i w_i^a w_i^b\right)\right\}\\
&\times& \prod_{t}\exp\left\{-\frac{1}{2}\sum_{a,b}\left(u_{t}^a+\epsilon^a-\mu_{t}^a\right)Q_{a,b}^{-1}\left(u_{t}^b+\epsilon^b-\mu_{t}^b\right)\right\}\exp\left\{-\frac{T}{2}{\rm tr}\log Q\right\}.
\end{eqnarray*}
Introducing the variables $y_{t}^a=\mu_{t}^a-u_{t}^b$ and
$z_{t}^a=\mu_{t}^a+u_{t}^b$ and integrating over the $\{z_{t}^a\}$ we obtain

\begin{eqnarray*}
Z_{\gamma}^n&=&  \int_{-i\infty}^{i\infty}\left(\prod_{a=1}^n\prod_{b=1}^n dQ_{a,b}d\hat Q_{a,b}\right)\int_{-\infty}^{\infty} \left(\prod_{a=1}^n d\epsilon^a\right)
\int_{-\infty}^{\infty}\left( \prod_{i=1}^N \prod_{a=1}^n d
w_i^a\right) \int_{-i\infty}^{i\infty}\left(\prod_{a=1}^n d\hat\lambda^a\right)\\
&\times& \exp\left\{\sum_a
\hat\lambda^a(\sum_i w_i^a-WN)\right\} \exp\left\{-\gamma\sum_a(1-\beta)T \epsilon^a-\gamma\sum_{a,t}
u_{t}^a-\gamma{\eta} \sum_i
{|w_i^a|^p}\right\}\\
&\times&\exp\left\{\sum_{a,b}\hat{Q}_{a,b}\left(N Q_{a,b}-\sum_i w_i^a w_i^b\right)\right\} \prod_{t}\exp\left\{-\frac{1}{2}\sum_{a,b}\left(u_{t}^a+\epsilon^a-\mu_{t}^a\right)Q_{a,b}^{-1}\left(u_{t}^b+\epsilon^b-\mu_{t}^b\right)\right\}\\
 &\times& \exp\left\{-\frac{T}{2}{\rm tr}\log Q-TN\log \gamma +T\log Z_{\gamma}(\{\epsilon^a,Q\})\right\}\end{eqnarray*}
where
\begin{eqnarray*}
Z_{\gamma}(\{\epsilon^a,Q\})&=&\int_{-\infty}^{+\infty} \prod_a dy^a \exp\left\{-\frac{1}{2}\sum_{a,b}(y^a-\epsilon^a)Q_{a,b}^{-1}(y^b-\epsilon^b)\right\}\\
&\times& \exp\left\{\gamma\sum_a y^a\theta(-y^a)\right\}.
\end{eqnarray*}

In order to make further progress, let us consider the replica symmetric ansatz
\begin{equation}
    Q_{a,b}= \left\{ \begin{array}{cc} q_1 ,&    a =b\\
    q_0 , &  a\neq b \end{array} \right.
\end{equation}
\begin{equation}
    \hat{Q}_{a,b}= \left\{ \begin{array}{cc} r_1 ,&    a =b  \\
    r_0 , &  a\neq b . \end{array} \right.
\end{equation}
and introduce the following rescaling relations
\bea
 \Delta &=&\gamma(q_1-q_0),\\
 \hat{\Delta}&=&(r_1-r_0)/\gamma,\\
 \lambda^a&=&\hat{\lambda}^a\gamma,\\
 \hat{q}_0&=&r_0\gamma^2.
\eea

The $\vec{w}$-dependent part of the partition function is
\bea
\int [D w] e^{-\gamma F_w} =\int [D w]  e^{\sum_{i a}
\lambda^a w_i^a-\gamma{\eta} \sum_i
{|w_i^a|^p}-\sum_{a,b}\hat{Q}_{a,b}\sum_i w_i^a w_i^b}\qquad.
\eea
Exploiting the identity $\log\langle X^n\rangle\simeq n\langle \log X\rangle$ valid for $n\to 0$,
and after some manipulations, we arrive at the following contribution to the free energy
\be
F_w={\gamma}\Big\langle \log\int dw e^{-\gamma \left[ \hat{\Delta} w^2+\eta |w|^p-{\lambda} w-z w\sqrt{-2{\hat{q}}_0}\right]}\Big\rangle_z,\qquad
\ee
where the notation $\langle\cdots\rangle_z$ means averaging over the normal variable $z$.
After some further manipulations, we can write the partition function as
\be\label{Zn}
Z_\gamma^n=\int d\lambda d \epsilon d q_0 d\Delta d{\hat{q}}_0 d\hat \Delta e^{-\gamma n N F[\lambda,{\epsilon},{q}_0,\Delta, {\hat{q}}_0,\hat{\Delta})]}
\ee
where
\bea
F( \lambda,{\epsilon},{q}_0,\Delta,  {\hat{q}}_0,\hat{\Delta})&=&
\lambda W +\tau (1-\beta)\epsilon -\Delta {\hat{q}}_0-\hat{\Delta} {q}_0\\
\nonumber &-&\frac{1}{\gamma}\Big\langle\log \int_{-\infty}^{\infty} dw e^{-\gamma V(w,z)}\Big\rangle_z
 +\frac{\tau\Delta}{2\sqrt{\pi}}\int_{-\infty}^{\infty}ds e^{-s^2}
g\left(\frac{\epsilon}{\Delta}+s \sqrt{2 \frac{{q}_0}{\Delta^2}}\right),
\eea

with $\tau=T/N$,
\be
V(w,z)=\hat{\Delta} w^2+\eta|w|^p-\lambda w -z w\sqrt{-2{\hat{q}}_0}
\ee
 and
\begin{equation}
    g(x)= \left\{ \begin{array}{cc} 0 ,&    x\ge 0\\
    x^2 , &  -1\le x\le 0\\
    -2 x-1, & x<-1
     \end{array} \right..
\end{equation}
In the limit of large $N$ the integral in equation \eqref{Zn} is concentrated around the minimum of $F( \lambda,{\epsilon},{q}_0,\Delta,  {\hat{q}}_0,\hat{\Delta})$ and  can be computed through the saddle point method.

The minimum cost 
\be
\lim_{\gamma\to\infty}\lim_{n\to 0}\left(-\frac{1}{\gamma}\frac{\partial Z_\gamma^n}{\partial n}\right)
\ee
corresponds then to the minimum of the free energy functional $F( \lambda,{\epsilon},{q}_0,\Delta,  {\hat{q}}_0,\hat{\Delta})$ in the limit $\gamma\to\infty$.

\bibliography{RPO}

\end{document}